# Using a multimode fiber as a high resolution, low loss spectrometer

Brandon Redding and Hui Cao[*]

*Department of Applied Physics, Yale University, New Haven, CT 06520*
*Corresponding author: hui.cao@yale.edu*

We propose and demonstrate that a conventional multimode fiber can function as a high resolution, low loss spectrometer. The proposed spectrometer consists only of the fiber and a camera that images the speckle pattern generated by interference among the fiber modes. While this speckle pattern is detrimental to many applications, it encodes information about the spectral content of the input signal which can be recovered using calibration data. We achieve a spectral resolution of 0.15 nm over 25 nm bandwidth using 1 meter long fiber, and 0.03 nm resolution over 5 nm bandwidth with a 5 meter fiber. The insertion loss is less than 10%, and the signal to noise ratio in the reconstructed spectra is over 1000.

Traditional spectrometers operate by mapping input signals of different wavelength to different spatial locations. In most implementations, signals within a spectral band are mapped to a specific area where a detector is placed to measure its intensity. While such one-to-one spectral-spatial mapping is necessary for a wavelength demultiplexer, it is not required for a spectrometer. In fact, spectrometers have been demonstrated which map a given spectral input to a complex spatial distribution of intensity [1, 2]. As long as distinct spatial patterns are produced by light at different wavelengths, an arbitrary input spectrum may be recovered from the calibration data. This approach allows the grating in a traditional spectrometer to be replaced by almost any dispersive element, e.g., an array of Bragg fibers [3], a disordered photonic crystal lattice [1], or even a random medium [2]. These spectrometers, however, afford only modest spectral resolution, while suffering high insertion loss and/or low signal to noise ratio (SNR).

In this letter, we demonstrate that a conventional multimode fiber can act as the dispersive element, enabling spectrometer operation using only a fiber and a camera. Interference of light propagating in multiple waveguided modes produces speckle, which varies with wavelength. Long propagation distance in the fiber leads to a rapid decorrelation of the speckle pattern with wavelength, giving high spatial-spectral diversity. After calibrating the speckle pattern as a function of wavelength, we reconstruct the spectra of input signals from the output speckles using a matrix pseudo-inversion algorithm in combination with a nonlinear optimization procedure. We achieve a spectral resolution of 0.15 nm over 25 nm bandwidth using a 1 m long fiber, and 0.03 nm resolution over 5 nm bandwidth with a 5 m fiber. The SNR is over 1000, and the insertion loss is less than 10%. Furthermore, the fiber can be coiled to provide a compact, light-weight, low-cost spectrometer which could enable a host of new spectroscopic applications.

Large core optical fibers can easily support hundreds of propagating modes, each having a different phase velocity. From a geometrical-optics point of view, various rays propagate down the fiber at different angles relative to the axis of the guide, thus they travel varying distances and experience different phase delays as they pass from the input to the output of the guide. For monochromatic input light, the electric field at any point on the exit face of the fiber consists of a sum of a multitude of individual field contributions:

$$\mathbf{E}(r,\theta,\lambda,L) = \sum_m A_m \mathbf{\psi_m}(r,\theta,\lambda)\exp\left[-i\left(\beta_m(\lambda)L - \omega t\right)\right], \quad (1)$$

where $A_m$ is the amplitude of the $m^{th}$ guided mode with spatial profile $\mathbf{\psi_m}$ and propagation constant $\beta_m$, $\lambda$ is the wavelength of light, and $L$ is the length of the fiber. The phase delay of a given mode is $\phi_m=\beta_m L$. After propagating over a long distance, the accumulated phase delays for different modes vary by more than $2\pi$ radians, thus speckle will be seen in the intensity distribution. While the speckle limits the use of multimode fibers in many applications, it contains information about the spectral content of the input signal. A shift in wavelength causes a change in phase delay $\Delta\phi_m=\Delta\beta_m L$ for the $m^{th}$ mode, where $\Delta\beta_m = \beta_m(\lambda+\Delta\lambda)-\beta_m(\lambda)$, and $\Delta\lambda$ is the wavelength shift. Although the change of propagation constant may be very small for a tiny wavelength shift, optical fibers can easily have length $L$ on the order of meters or longer, making $\Delta\phi_m$ comparable to $\pi$ radians and causing a dramatic change in the speckle pattern. In other words, the speckle image provides a fingerprint of the input wavelength for a multimode fiber of fixed parameters.

To enable high spectral resolution, the speckle must decorrelate quickly with wavelength. To test this, we experimentally measured the speckle patterns at the end of multimode fibers as a function of wavelength. We used commercially available step-index multimode fiber patch cables with length of 1 m, 2 m and 5 m. The fiber core diameter is 105 μm, and the numerical aperture (NA) is 0.22. These fibers support ~1000 modes at $\lambda$ = 1500 nm [4]. A tunable diode laser (Hewlett Packard 8168F) provided a spectrally controlled input signal through a polarization-maintaining single mode fiber, which was coupled to the multimode fiber via a standard FC/PC connector. The speckle pattern generated at the exit face of the multimode fiber was then imaged onto an InGaAs camera (Xenics Xeva 1.7-320) with a 50× objective lens (NA = 0.55). Example speckle patterns collected at three closely spaced wavelengths with the 5 m long fiber are presented in Fig. 1(a-c). The measured speckle pattern changes significantly for wavelengths separated by merely 0.02 nm, indicating extremely high sensitivity to the input wavelength. By recording speckle images at different wavelengths, we were able to calculate the spectral correlation function $C(\Delta\lambda) = \langle I(\lambda)I(\lambda+\Delta\lambda)\rangle / [\langle I(\lambda)\rangle\langle I(\lambda+\Delta\lambda)\rangle] - 1$, where $I(\lambda)$ is the intensity of light at a given location and wavelength $\lambda$, and $\langle \cdots \rangle$ represents the average over $\lambda$. In Fig. 1(d), we plot the spectral correlation function averaged over many spatial positions across the fiber core. The speckle pattern decorrelates more quickly in longer fibers. In Fig. 1(e), we show that the spectral correlation width $\delta\lambda$ [equal to twice of the half width at half maximum of $C(\Delta\lambda)$] scales as the reciprocal of the length of the fiber.

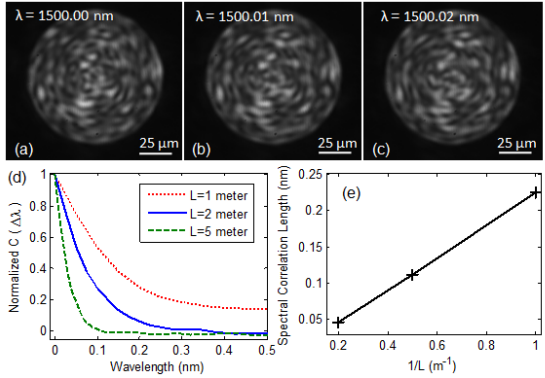

Fig. 1. (Color online) (a-c) Images of the intensity distribution at the end of a 5 m long multimode fiber with varying input wavelength. While the speckle pattern in (a) shows some resemblance to that in (b), it looks very different from that in (c), illustrating that the speckle pattern decorrelates quickly with wavelength. (d) Spectral correlation function $C(\Delta\lambda)$ normalized to unity at $\Delta\lambda=0$ for fibers of varying length. (e) The spectral correlation width $\delta\lambda$, as a function of the inverse of the fiber length $L$. The crosses represent the experimental data points, and the straight line is a linear fit showing that $\delta\lambda$ scales as $1/L$.

In order to use the multimode fiber as a spectrometer, we must first calibrate the transmission through the fiber. The intensity at the output end of a fiber can be written as

$$I(r,\theta) = \int T(r,\theta,\lambda) S(\lambda) d\lambda, \quad (2)$$

where $S(\lambda)$ is the illuminant spectral flux density at the input end of the fiber, and $T(r,\theta,\lambda)$ is the position-dependent transmission function. Although they can be arbitrary, the spatial profile and polarization state of light at the input end must be fixed, otherwise the transmission function will change. Spectral discretization produces spectral channels centered at $\lambda_i$ and spaced by $d\lambda$. If $d\lambda=\delta\lambda$, the spectral channels are independent. Similarly, spatial discretization over the cross section of the fiber core generates spatial channels centered at $r_j$, and they become independent when their spacing exceeds the spatial correlation length, $l_c$. After discretization, Eq. (2) becomes $I_j = T_{ji} S_i$. $T_{ji}$ forms an $M \times N$ matrix, where $M$ and $N$ are the number of spatial and spectral channels, respectively. The transmission matrix $T$ can thus be calibrated one column at a time by recording the output intensity distribution after tuning the input laser wavelength to $\lambda_i$ (the laser linewidth is much smaller than $d\lambda$). Note that intermodal coupling caused by bending and twisting of the fiber is allowed and has been accounted for by the transmission matrix, but it must remain unchanged after the calibration.

In principle, the input spectra may be recovered from the output speckle by inversion of the transmission matrix. In the case of independent spectral and spatial channels, $T$ can be inverted if $M=N$. However, experimental noise makes the inversion process ill-conditioned. To reduce the effects of noise, we performed oversampling by choosing the spacing of spectral channels $d\lambda=\delta\lambda/4$ and setting the distance of spatial channels at $l_c/2$. In Fig. 2(a), we show a close-up of a typical speckle image where the locations of spatial channels are marked with "+" symbols. For each fiber, we selected 500 spatial channels and 500 spectral channels. Figure 2(b) is part of the transmission matrix $T$ for the 1 m long fiber. To reconstruct the original spectra, we adopted a matrix pseudo-inversion algorithm based on singular value decomposition [3]. The singular values below a threshold are truncated to optimize the SNR in the reconstructed spectra. To further reduce the reconstruction error, we then performed a nonlinear optimization procedure to find the spectra $S_i$ that minimizes $\Sigma_j |I_j - \Sigma_i T_{ji} S_i|^2$ [1].

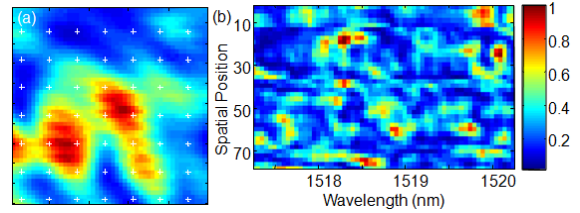

Fig. 2. (Color online) (a) Close-up of a speckle pattern imaged at the end of a 1 m fiber. The transmission matrix was sampled at spatial positions separated by 3.8 μm ($l_c/2$), indicated by "+" symbols. (b) Part of the transmission matrix for a 1 m fiber describing the intensity generated at varying spatial positions by different wavelengths. The wavelength spacing of adjacent spectral channels is 0.05 nm ($\delta\lambda/4$).

To test the performance of the fiber spectrometer, we set the input wavelength in between the sampled wavelengths $\lambda_i$ used to calibrate the transmission matrix, and measured the output speckle. We reconstructed spectra for 25 separate input wavelengths using the 1 m fiber (25 nm bandwidth with 500 spectral channels spaced by $d\lambda$=0.05 nm), as shown in Fig. 3(a). The average reconstructed spectral linewidth is about 0.12 nm and the average SNR is greater than 1000. In Fig. 3(b), we show that the 1 m fiber spectrometer can clearly resolve two spectral lines separated by 0.15 nm. This test was conducted by measuring the speckle patterns at two wavelengths separately then adding their intensities because light at different wavelengths does not interfere. The spectral resolution is slightly larger than the individual linewidth due to reconstruction noise. Figure 3(c-d) presents the results of characterization of the 5 m fiber spectrometer. The spectral resolution is 0.03 nm over a 5 nm bandwidth, and the SNR is greater than 1000.

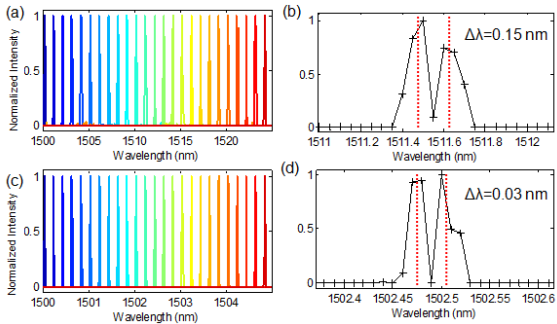

Fig 3. (Color online) (a, c) Reconstructed spectra of single lines centered between the wavelengths sampled for the *T* matrix calibration. The average linewidth for the 1 m fiber spectrometer (a, 25 nm bandwidth, 0.05 nm channel spacing) is 0.12 nm, and for the 5 m fiber (c, 5 nm bandwidth, 0.01 nm channel spacing) is 0.021 nm. (b, d) Reconstructed spectra (black solid lines with crosses marking the sampled wavelengths $\lambda_i$) of two closely spaced lines. Red dotted lines mark the center wavelengths of the input lines. The 1 m fiber spectrometer (b) can clearly resolve two lines separated by 0.15 nm, and the 5 m fiber (d) can resolve lines separated by 0.03 nm.

In Table 1, we summarize the performance of three spectrometers with fibers of length $L$=1 m, 2 m, and 5 m. The spectral resolution scales inversely with $L$. For the same number of spectral channels ($N$=500), the operation bandwidth also scales as $1/L$. In addition to providing high spectral resolution and broad operation bandwidth, the multimode fiber spectrometer has extremely low loss. We measured an insertion loss of less than 10% by comparing the input intensity of the laser coming out of the single mode fiber and the output intensity from the end of the multimode fiber. This loss is dominated by the coupling between the single mode fiber and the multimode fiber which could be optimized for higher efficiency.

**Table 1. Summary of spectrometer performance**

| Fiber Length | Spectral Correlation | Channel Spacing | Spectral Resolution | Operation Bandwidth |
|---|---|---|---|---|
| 1 m | 0.224 nm | 0.05 nm | 0.15 nm | 25 nm |
| 2 m | 0.112 nm | 0.025 nm | 0.075 nm | 12.5 nm |
| 5 m | 0.046 nm | 0.01 nm | 0.03 nm | 5 nm |

Not only does the fiber spectrometer provide competitive resolution, bandwidth and insertion loss with larger, more expensive spectrometers, but the transmission matrix technique also enables some unique functionalities. As in most spectrometers, the spatial distribution of the input signal affects the measured spectra. For the fiber spectrometer, the spatial distribution and polarization of the input signals must be identical to those used in the calibration. This is guaranteed in our implementation by using a polarization maintaining single mode fiber to provide a fixed input to the multimode fiber. However, a single mode fiber is not strictly required, and in many sensing applications it is advantageous to detect diffuse light without the spatial filtering imposed by a single mode fiber (or the entrance slit of a traditional spectrometer) [5]. We expect the multimode fiber spectrometer can also operate with diffuse input, provided the calibration is performed using the same spatial input.

Another concern for any spectrometer is the elimination of stray light. For the multimode fiber, optical signals outside the spectral bandwidth of operation are not included in the calibrated transmission matrix, and their presence will contribute to reconstruction error. While care must be taken to filter signals outside the operation bandwidth, this also highlights a potential advantage of the fiber spectrometer. A single multimode fiber can operate at varying spectral regions similar to the way traditional spectrometers do by rotating the grating. For a fiber spectrometer, the analog of rotating the grating is achieved by switching the transmission matrix to one calibrated for the spectral region of interest.

While conventional spectrometers map spectral information to one spatial dimension, the fiber spectrometer maps to two-dimensions (2D). This 2D spatial-spectral mapping fully utilizes the large detection area of modern 2D cameras to achieve both high spectral resolution and large bandwidth of operation. The maximal number of independent spectral channels that can be measured in parallel is limited by the number of independent spatial channels (i.e. number of speckle correlation cells), which is equal to the number of propagating modes in the fiber if all modes are more or less equally excited. Like a conventional spectrometer, there is a trade-off between the spectral resolution and the operation bandwidth for the fiber spectrometer.

In summary, we demonstrated that a high resolution, low loss spectrometer can be implemented in a multimode fiber with a 2D camera. Our approach is applicable to any wavelength range. The long propagation length of light in the fiber results in high spectral resolution, and the large core diameter enables broadband operation. Crucially, the fiber length and core diameter can be increased without significantly affecting the insertion loss of the spectrometer. Compared to existing spectrometers with comparable resolution and bandwidth, the multimode fiber spectrometer is compact, lightweight and inexpensive. Furthermore, our technique lends itself to the development of a hyperspectral imaging system using a fiber bundle with multimode cores. Each core acts as a spectrometer with its own transmission matrix, probing the spectra of light from a local area. By imaging the speckle patterns generated within individual cores, the spectral contents of signals associated with different spatial locations can be recovered simultaneously.

We acknowledge stimulating discussion with Dr. Sebastien Popoff. This project is supported partly by NSF under the Grant No. ECCS-1128542.

### References
1. Z. Xu, Z. Wang, M. E. Sullivan, D. J. Brady, S. H. Foulger, and A. Adibi, Opt. Expr. **11**, 2126 (2003).
2. T. Kohlgraf-Owens, and A. Dogariu, Opt. Lett. **35**, 2236 (2010).
3. Q. Hang, B. Ung, I. Syed, N. Guo, and M. Skorobogatiy, Appl. Opt. **49**, 4791 (2010).
4. J. W. Goodman, *Speckle phenomena in optics,* (Roberts & Company, 2007).
5. O. Momtahan, C. R. Hsieh, A. Karbaschi, A. Adibi, M. E. Sullivan, and D. J. Brady, Appl. Opt. **43,** 6557 (2004).